\documentclass[a4paper,11pt]{article}
\pdfoutput=1
\usepackage{jinstpub}
\usepackage{multirow}
\usepackage{array}
\usepackage{lineno}
\newcommand{\PreserveBackslash}[1]{\let\temp=\\#1\let\\=\temp}
\newcolumntype{C}[1]{>{\PreserveBackslash\centering}p{#1}}

\author[a]{H. Pernegger \footnote{Corresponding author, \tt{Email: heinz.pernegger@cern.ch}}}
\author[b]{P. Allport}
\author[c]{D.V. Berlea}
\author[l]{A. Birman}
\author[d]{D. Bortoletto}
\author[e]{C. Buttar}
\author[f]{E. Charbon}
\author[a]{F. Dachs}
\author[a]{V. Dao}
\author[g]{H. Denizli}
\author[a,h]{D. Dobrijevic}
\author[a]{M. Dyndal \footnote{now at AGH UST Krakow, Poland}}
\author[l]{A. Fenigstein}
\author[a]{L. Flores Sanz de Acedo}
\author[b]{P. Freeman}
\author[a]{A. Gabrielli}
\author[d]{M. Gazi}
\author[b]{L. Gonella}
\author[i]{V. Gonzalez}
\author[a]{G. Gustavino}
\author[k]{A. Haim}
\author[a]{T. Kugathasan}
\author[a]{M. LeBlanc}
\author[a]{M. Munker \footnote{now at University of Geneva, Geneva, Switzerland}}
\author[g]{K.Y. Oyulmaz}
\author[a,f]{F. Piro}
\author[a]{P. Riedler}
\author[j]{H. Sandaker}
\author[a]{E.J. Schioppa \footnote{now at Universit\`a del Salento, Lecce, Italy}}
\author[a]{A. Sharma}
\author[a]{W. Snoeys}
\author[a]{C. Solans Sanchez}
\author[h]{T. Suligoj}
\author[k]{E. Toledano}
\author[a,j]{M. van Rijnbach}
\author[a,i]{M. Vazquez Nunez}
\author[a,m]{J. Weick}
\author[c]{S. Worm}
\author[m]{A.M. Zoubir}

\affiliation[a]{CERN, Geneva, Switzerland}
\affiliation[b]{University of Birmingham, Birmingham, UK}
\affiliation[c]{DESY, Zeuthen, Germany}
\affiliation[d]{University of Oxford, Oxford, UK}
\affiliation[e]{University of Glasgow, Glasgow, UK}
\affiliation[f]{EPFL, Lausanne, Schwitzerland}
\affiliation[g]{Bolu Abant Izzet Bazsal, Merkez/Bolu,Turkey}
\affiliation[h]{University Zagreb, Zagreb, Croatia}
\affiliation[i]{Universitat de Val\`encia, Val\`encia, Spain}
\affiliation[j]{University of Oslo, Oslo, Norway}
\affiliation[k]{Etesian Semiconductor, Ramat Yishai, Israel}
\affiliation[l]{Tower Semiconductor, Migdal Haemek, Israel}
\affiliation[m]{Technische Universit\"at Darmstadt, Darmstadt, Germany}

\title{MALTA-Cz: A radiation hard full-size monolithic CMOS sensor with small electrodes on high-resistivity Czochralski substrate}

\abstract{Depleted Monolithic Active Pixel Sensor (DMAPS) sensors developed in the Tower Semiconductor 180 nm CMOS imaging process have been designed in the context of the ATLAS ITk upgrade Phase-II at the HL-LHC and for future collider experiments. The ``MALTA-Czochralski (MALTA-Cz)'' full size DMAPS sensor has been developed with the goal to demonstrate a radiation hard, thin CMOS sensor with high granularity, high hit-rate capability, fast response time and superior radiation tolerance. The design targets radiation hardness of $>10^{15}$ (1~MeV) n$_{eq}$/cm$^{2}$ and 100~Mrad TID. The sensor shall operate as tracking sensor with a spatial resolution of $\approx$10~~$\mu$m and be able to cope with hit rates in excess of 100~MHz/cm${^2}$ at the LHC bunch crossing frequency of 40MHz. The 512$\times$512 pixel sensor uses small collection electrodes (3.5~$\mu$m) to minimize capacitance. The small pixel size ($36.4\times 36.4$~$\mu$m$^2$) provides high spatial resolution. Its asynchronous readout architecture is designed for high hit-rates and fast time response in triggered and trigger-less detector applications. The readout architecture is designed to stream all hit data to the multi-channel output which allows an off-sensor trigger formation and the use of hit-time information for event tagging.  

The sensor manufacturing has been optimised through process adaptation and special implant designs to allow the manufacturing of small electrode DMAPS on thick high-resistivity p-type Czochralski substrate. The special processing ensures excellent charge collection and charge particle detection efficiency even after a high level of radiation. Furthermore the special implant design and use of a Czochralski substrate improves the sensor's time resolution. This paper presents a summary of sensor design optimisation through process and implant choices and TCAD simulation to model the signal response. Beam and laboratory test results on unirradiated and irradiated sensors have shown excellent detection efficiency after a dose of  $2\times10^{15}$ 1 MeV n$_{eq}$/cm$^{2}$.  The time resolution of the sensor is measured to be $\sigma=2$~ns.}

\keywords{Radiation-hard detectors, Solid state detectors, CMOS imagers, Monolithic Active Pixel Sensor, Detector modelling and simulations}

\date{\today}

\notoc

\begin{document}

\maketitle

\section{Introduction}

Depleted Monolithic Active Pixel Sensor (DMAPS) prototypes have been developed in the Tower Semiconductor 180~nm CMOS imaging process with the aim to explore their use in the Phase-II upgrade of ATLAS for the High Luminosity LHC ~\cite{ItkPixelTdr}, and for future HEP experiments \cite{Investigator}\cite{Snoeys:2017hjn}. Monolithic CMOS sensors allow the minimization of scattering material for best tracking performance. Furthermore they reduce construction costs and assembly time of detector systems due to absence of bump-bonding required in hybrid sensors. With previous developments focusing on low-radiation environments \cite{Algieri:2013}, special interest lies now on the radiation hardness of this technology up to 100 Mrad in Total Ionizing Dose (TID) and $\ge 1\times10^{15}$ 1 MeV n$_{eq}$/cm$^{2}$ in Non-Ionizing Energy Loss (NIEL) in order to be used in the harsh environment of pp-collider experiments at LHC and future colliders. 

The developments reported here investigate pixel sensors based on the MALTA sensor architecture \cite{Cardella:2019ksc}\cite{Berdalovic-2018}. Its pixel design and sensor processing is specially chosen to achieve radiation hardness while maintaining the advantages of pixels with small electrodes:  The small electrode (3.5~$\mu$m diameter) results in small capacitance, which in turn helps to minimize noise and achieve low power dissipation in the active area. 
Traditionally sensors with small electrode suffer from reduced detection efficiency after irradiation when charge collection is critically affected in the pixel corners \cite{Cardella:2019ksc}\cite{Caicedo2019}\cite{Wang2018} by radiation induced charge trapping.  We designed special p-type and n-type implant geometries \cite{Munker:2019vdo} to improve charge collection in pixel corners. Prototype sensors produced on high-resistivity epitaxial substrate have shown significantly improved corner efficiency when using these special pixel implant design and higher-gain front-end electronics \cite{dyndal2019}. 

We have implemented MALTA's novel pixel designs for the first time on thick high-resistivity Czochralski substrates: The manufacturing of small electrode CMOS sensors with special implant geometries on high-resistivity Czochralski substrates combines the advantages of small electrode sensors with the advantages of a thicker detection layer: The low pixel capacitance is maintained for low noise and low power performance. At the same time the signal amplitude is significantly increased due to the larger ionisation charge in thicker depleted sensors. For direct soft X-ray sensing application furthermore the quantum efficiency significantly increases by enlarging the detection volume thickness from $\approx$25~$\mu$m of epitaxial silicon layer to $\approx$100~$\mu$m or larger depending on depletion voltage and resistivity. Implementing the identical sensor design on high-resistivity substrates allows the optimization of sensors for multiple applications simultaneously, to serve high energy and nuclear physics experiments as well as other industrial and research applications like cryogenic electron microscopy.

\section{The MALTA sensor design}

The small collection electrode minimizes input capacitance to achieve a high $Q/C$ ratio in the front-end amplifier. This results in a large voltage swing in response to a minimum ionising particle in the pre-amplifier which facilitates the circuit and improves signal-to-noise. In the case of a 25~$\mu$m thick epitaxial detection layer we expect a most probable ionization charge of around 1500~e$^{-}$. To calculate the expected ionisation charge for thin sensors, we assume an ionisation charge of 63 electron-hole pairs per $\mu$m path length~\cite{Meroli}. Realising the same sensor design on high-resistivity Czochralski substrates enables us to increase the signal amplitude by a large factor. In case of e.g. 100~$\mu$m thick depletion layer, the signal increases to more than 6000~e$^{-}$.
With a total electrode capacitance of 5~fF this deposited charge causes a voltage step of around 50 mV on epitaxial substrate and over 200~mV on Czochralski substrates. 

The large voltage swing offers the possibility of using an open-loop voltage amplifier as the first amplification stage for the MALTA-Cz sensor, instead of the conventional charge-sensitive amplifier scheme with a feedback capacitor. The open-loop amplifier simplifies the front-end circuit in the pixel and saves space. The input node is reset after a particle hit using a diode-circuit. The front-end amplifier output connects to a discriminator, which produces the digital output signal of the pixel. The discriminator threshold is set globally for the full sensor.  The analog front-end circuit is described in more detail in \cite{Cardella:2019ksc,Berdalovic-2018}. The front-end circuit is designed to operate at a threshold of $\approx$200~e$^{-}$ with a sufficiently fast response for the 25~ns timing requirement of the HL-LHC bunch crossing. The small electrode capacitance allows to operate the front-end circuit with a bias current of 500~nA per pixel for this timing requirement, which gives an analog power density of 75~mW/cm$^{2}$. The analog circuit includes a clipping mechanism to achieve a signal return to baseline after 200~ns, which reduces dead time for high-rate applications. Each pixel also includes a charge injection test capacitance which allows to pulse a single pixel with a rectangular voltage step of programmable amplitude. The test capacitance is formed by a parasitic capacitance between two metal lines which has been extracted from simulation as approximately 230~aF. This charge injection is used to determine the effective threshold on a pixel through a so-called threshold-scan: ``S-curves'' are measured as hit occupancy versus injected charge. The S-curve is fitted with a complementary error function from which the 50\%-occupancy point of the S-curve gives the pixel threshold in units of charge and the $\sigma$ of the complementary error function gives the pixel noise as equivalent noise charge.  

\begin{figure}
    \centering
    \includegraphics[width=0.7\textwidth]{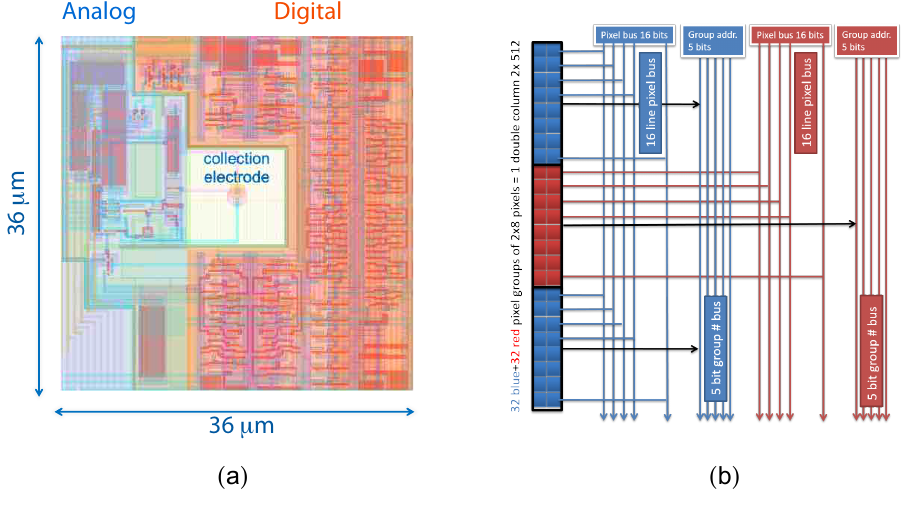}
     \caption{The MALTA pixel design with small charge collection electrode (a) and asynchronous readout architecture (b) \cite{Cardella:2019ksc,Berdalovic-2018,PERNEGGER2021164381,RIEDLER2021164895}.
    }
    \label{fig:MALTA}
\end{figure}

The full-size MALTA-Cz sensor is comprised of the 512$\times$512 pixel matrix with an active area of 18.3$\times$18.3 mm$^{2}$ ($36.4\times 36.4$~$\mu$m$^2$ pixel pitch) \cite{PERNEGGER2021164381,RIEDLER2021164895}. The charge collecting electrode sits in the center of the pixel as shown in figure~\ref{fig:MALTA}a . The hit signals are transmitted asynchronously from the pixel discriminator to the sensor periphery along the column. The design avoids the distribution of high frequency clock signals across the matrix to conserve power and minimise analog-digital crosstalk. To cope with high hit-rates well above 100~MHz/cm${^2}$ we have implemented a dedicated group-logic in the double-column: Pixels are organised in groups of 2$\times$8 pixels as shown in figure \ref{fig:MALTA}b. Each pixel output in a group is routed to one of 16 lines in a so-called ``pixel'' bus. This 16-bit pixel-bus is shared between every other 2$\times$8-group (e.g. shown as blue groups in figure \ref{fig:MALTA}b). To identify groups in a double column we use a 5-bit ``group'' address bus. The double column readout architecture is formed by interleaving two identical buses, denoted as ``blue'' and ``red'' groups. When a pixel discriminator switches to ON, a programmable width signal (0.5 to 2ns signal width) is generated as reference signal (``$V_{ref}$'') on one line of the double column bus and simultaneous signals on the 16-bit pixel-bus and 5-bit group-bus denote the hit pixel address. 

This architecture and grouping has been invented to address several conditions simultaneously: For a high-density pixel detector with small pixel pitch and large column height of nearly 2cm, the pixel area dedicated to output line routing is space limited. The bus sharing between groups allows to minimise the number of address lines in a 512-row double column to 44 lines. At the same time sensors for high-hit rate applications have to cope with multiple particle hits in a double column in a single bunch crossing cycle of 25~ns. The combination of short digital pulse width with interleaved groups on separated buses allows to operate this architecture at high hit rates without significant digital inefficiency in the matrix due to decoding errors on the hit address bus ($<$1\% at 1~GHz/cm$^{2}$ and $<$0.05\% at  200~MHz/cm$^{2}$). 

\section{Sensor manufacturing and pixel implant designs} \label{process}

\begin{figure}
    \centering
    \includegraphics[width=0.95\textwidth]{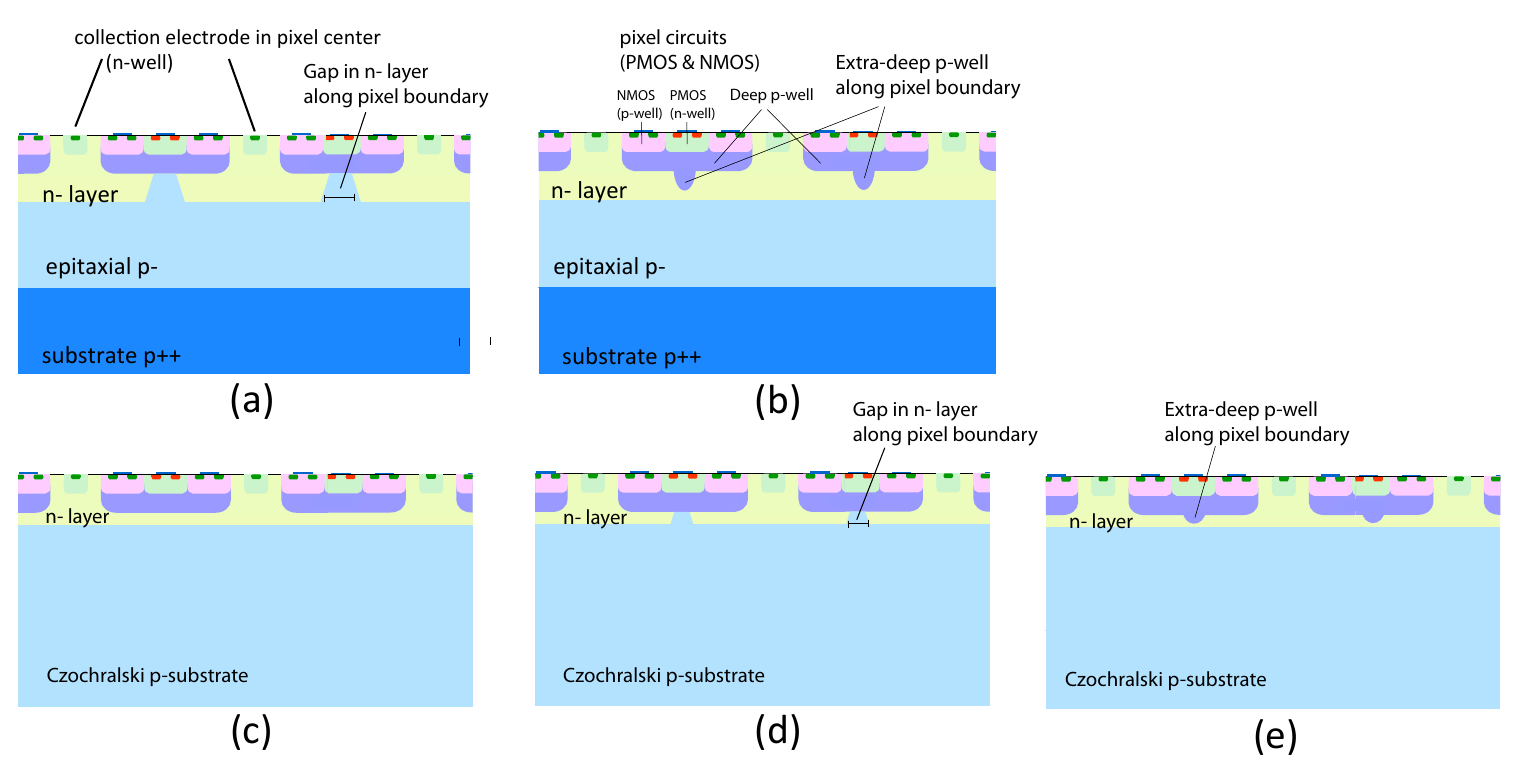}

    \caption{Cross section of the Tower process for the MALTA-Czochralski sensor evaluation: Sensors with low dose n$^-$ layer in epitaxial layer including a n$^-$ gap (a) and extra deep p-well at the edge of the pixel (b). Sensors with a continuous low dose n$^-$ layer on Czochralski substrate (c),  with the low dose n-implant removed (n$^-$ gap) at the edge of the pixel (d) and with an extra deep p-well at the edge of the pixel (e).
    }
    \label{fig:process_modification}
\end{figure}

The pixel implant geometry used in the Tower Semiconductor 180~nm CMOS imaging process has been optimised for fast charge collection and high radiation hardness in collaboration with the foundry. Figure~\ref{fig:process_modification} shows the cross-sections of different implant designs and substrate combinations which have been produced for this study. The ionisation signal is collected on the small n-well collection electrode in the pixel center (3.5$\mu$m diameter). The electrode is connected to the analog front end, located together with the digital in-pixel circuitry in the surrounding deep p-well (deep p-well to allow n-wells for PMOS transistors in the circuit). Our tests are carried out on two sectors of the full matrix with different electrode configurations: On MALTA sector ``2'' the collection electrode is surrounded by the p-well and deep p-well  at a distance of 3.5~$\mu$m. On MALTA sector ``3'' the collection electrode is surrounded by the p-well at a distance of 3.5~$\mu$m however the deep p-well is further retracted to larger distance from the electrode within the limits of PMOS transistor placement \cite{Cardella:2019ksc}. 

The standard imaging process is supplemented with a low-dose $n$-type implant across the full pixel matrix~\cite{Snoeys:2017hjn}. This n$^{-}$ layer generates the depletion of the silicon sensor and separates the deep p-well of the pixel circuit from the p-type substrate. The substrate is reverse biased  to deplete the epitaxial layer or the high-resistivity Czochralski substrate bulk. The substrate bias is supplied in parallel from the top side through a substrate p-type connection as well as the sensor backside through an electrical connection on the chip-carrier PCB. The p-well is biased in our measurements at $-$6 V unless stated otherwise. The n$^{-}$ layer is depleted from its junctions to the deep-p-well on one side and p-type substrate on the other side. The choice of n$^{-}$ layer doping concentration, the distance between p-well and collection n-well and the size of collection electrode influences the detector capacitance seen by the front-end input and as a consequence its analog gain.  MALTA pixel designs have been optimised in TCAD and circuity simulations, and in measurements on prototype sensors. 

The manufacuring process was further modified by adding a 4~$\mu$m wide gap in the low dose n-layer along the pixel edges through a mask change (figure~\ref{fig:process_modification}a and ~\ref{fig:process_modification}d) or adding an additional production process compatible extra deep p-type implant of 4~$\mu$m width underneath the normal deep p-well (figure~\ref{fig:process_modification}b and \ref{fig:process_modification}e). We refer to these configurations as ``n$^{-}$ gap'' and ``extra deep p-well'' configurations, respectively. The purpose of these modifications was to improve the charge collection at the pixel edges and corners through the creation of a stronger lateral field, which focuses the ionization charge towards the collection electrode. The design of these implant structures has been optimized in TCAD simulations~\cite{Munker:2019vdo}.  Previous measurements on smaller-sized sensor arrays have shown that these measures can substantially increase detection efficiency in the corners after irradiation \cite{dyndal2019}. Furthermore these modifications increase the charge collection speed, resulting in a sensor with fast signal response as will be shown in section \ref{sec:sim}.

Figure~\ref{fig:process_modification}a and \ref{fig:process_modification}b show the implant configurations on epitaxial substrate with n$^{-}$ gap and extra deep p-well. The p-type epitaxial layer with a resistivity of $>$1000~$\Omega$cm has a thickness of 30~$\mu$m which results in an active sensor thickness of approximately 30~$\mu$m, depleted approximately over the full depth. Figure~\ref{fig:process_modification}c shows the implant configuration with a continuous n$^{-}$ layer across the sensor active area processed on $>$800~$\Omega$cm p-type Czochralski substrate \footnote{as given in substrate manufacturing specification}. Figure~\ref{fig:process_modification}d and \ref{fig:process_modification}e show the implant configurations on Czochralski substrate with n$^{-}$ gap and extra deep p-well. Spreading resistance profile measurements on completed sensors indicate a bulk resistivity of epitaxial as well as Czochralski substrates in excess of 3~k$\Omega$cm. The sensors were back-thinned to either 300~$\mu$m or 100~$\mu$m thicknesses, and no back-side implantation was carried out. 

To achieve sensors manufactured on Czochralski substrate with large active sensor volume (>50~$\mu$m depletion width) we require good electrical separation of (deep-)p-well and p-type substrate through the n$^{-}$ layer. Only in this case we can reverse bias the substrate with significantly higher substrate voltages than V$_{p-well}$, which is the limited to -6~V. Early punch-through between substrate and p-well would limit the depletion of of the Czochralski substrate. For this reason the geometrical dimensions of implants and doping profiles have been carefully optimised in collaboration with the foundry.

With non-ionising energy loss from hadron irradiation, the properties of the high resistivity sensor bulk (epitaxial or Czochralski) will change. Specifically the effective doping concentration $N_{eff}$ will change through the creation of deep level acceptor and donor traps. With heavily irradiated high-resistivity substrates the effective p-type doping of the high-resistivity bulk will increase and its resistivity will decrease \cite{Cindro2009,Mandic2017}. The change of $N_{eff}$ in p-type high-resistivity ($>$2~k$\Omega$cm) Czochralski substrate follows the fluence as $N_{eff} = g_{c} \Phi_{eq}$ at high fluences with $g_{c}$ ranging from $g_{c} = 0.022$ \cite{Cindro2009} to $g_{c} = 0.047$ for neutrons \cite{Mandic2017}. Assuming an initial resistivity of $>$2~k$\Omega$cm we can expect from data presented in \cite{Mandic2017} an effective p-doping concentration of $\approx 5 \times 10^{13}$cm$^{-3}$ ($\approx 300~\Omega$cm) after a neutron irradiation of $1\times10^{15}$ n$_{eq}$/cm$^{2}$ and $\approx 10^{14}$cm$^{-3}$ ($\approx$150~$\Omega$cm) after an irradiation of $2\times10^{15}$ n$_{eq}$/cm$^{2}$. Applying a reverse bias voltage of -50~V to the substrate yields a calculated depletion thickness of 110~$\mu$m for an unirradiated sensor of 3~k$\Omega$cm resistivity, 37~$\mu$m for a $1\times10^{15}$ n$_{eq}$/cm$^{2}$ irradiated sensor of 300~$\Omega$cm resistivity and 25~$\mu$m for a $2\times10^{15}$ n$_{eq}$/cm$^{2}$ irradiated sensor of 150~$\Omega$cm resistivity. It should be noted that the irradiated sensors used in this paper have been subjected to short beneficial annealing (several days at room temperature) but not to long-term reverse annealing at elevated temperatures. 

Another consequence of high neutron or proton irradiation to the sensor can be the creation of a so-called ``double junction'' through the accumulation of positive or negative space charge near p$^{+}$ or  n$^{+}$ wells \cite{Eremin2002}. Through non-ionizing irradiation, deep level defects are created in the bulk which induce acceptor and donor states that act as charge traps for free charge carriers. The generation (e.g. thermal or ionisation) hole current will accumulate at the p$^{+}$ contacts and the electron current at the n$^{+}$ contacts. The non-uniformity of currents then leads to a non-uniformity of space charge distribution: Near the p$^{+}$-well the predominant hole current will fill deep donor traps if we assume that the number of filled traps is proportional to the number of free carriers. This leads to a build-up of positive space charge near the p$^{+}$-well, which behaves like an effective n-doped region. Similar negative space charge is built up in the region near n-implants and the region close to the implant behaves like it is p-doped. While the ``double junction'' effect has primarily been studied for float-zone silicon, it  has also been measured through transient current techniques in n-type and p-type Czochralski silicon \cite{Pacifico2010}.

\section{Simulation of signal response in epitaxial and Czochralski sensors} \label{sec:sim}

Three-dimensional TCAD simulations have been previously employed to optimise the electrode configuration and implant geometries~\cite{Munker:2019vdo}. Similar simulations have been used to compare field configurations and transient current signal of the MALTA pixel produced on epitaxial and Czochralski substrate. For the purpose of simulation we assumed a 25$\mu$m thick epitaxial silicon layer and a 150$\mu$m thick Czochralski substrate which mimics average values of our produced sensors. The simulation has been performed with voltage configurations typically used in tests, i.e. 0.8~V on the collection electrode and -6~V bias on the p-well. The substrate voltage was varied in simulation to investigate punch-through effects between p-well and substrate in TCAD simulation. The simulation uses implantation profiles for n-well, p-well, deep p-well and the additional n$^{-}$ layer provided by the foundry from process simulation. NIEL radiation induced defects are modelled in TCAD as given in reference \cite{moscatelli2016} for the range up to  $7\times10^{15}$ n$_{eq}$/cm$^{2}$. It should be noted, that the acceptor and donor parameters (energy, cross section, introduction rate) in \cite{moscatelli2016} are optimised for p-type float zone silicon in proton irradiations, hence the TCAD simulation of radiation effects on depletion and charge trapping for our sensors should be considered as approximate.

\begin{figure}[!htb]
 \centering
    \includegraphics[width=.505\textwidth]{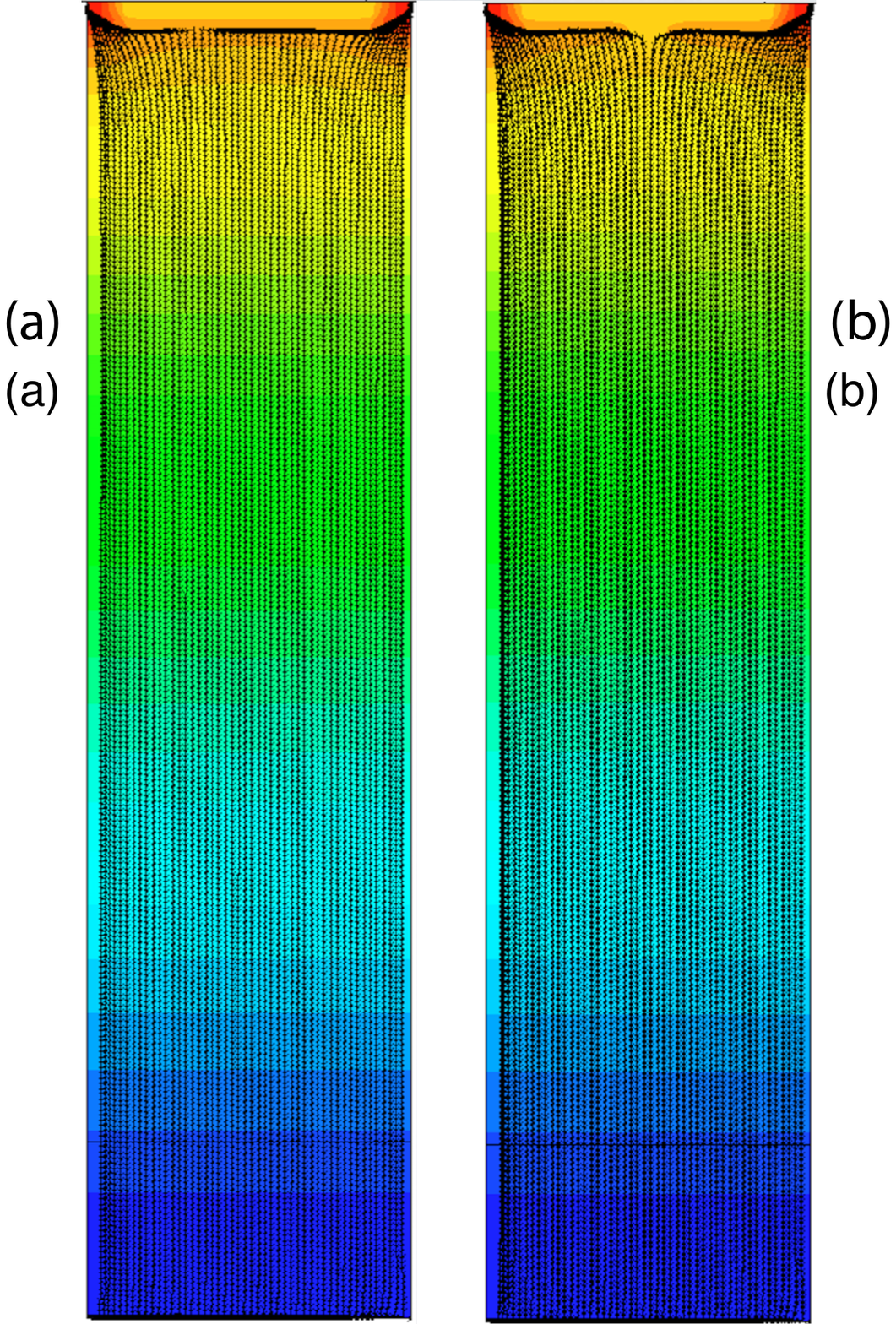}
    \caption{Electrostatic potential and electric field lines obtained from TCAD simulations of the Cz-sensor with continuous n$^-$ layer (a) (c.f. figure \ref{fig:process_modification}c) and of the Cz-sensor with extra deep p-well (b) (c.f. figure \ref{fig:process_modification}e)}
    \label{fig:tcad-field}
\end{figure}

Figure~\ref{fig:tcad-field} shows the simulated electrostatic potential (colour) and the electric field lines for a 150~$\mu$m thick 3~k$\Omega$cm Czochralski sensor with a substrate bias voltage of -50~V. Figure~\ref{fig:tcad-field}a assumes a continuous n$^-$ layer across the full pixel as is illustrated in figure \ref{fig:process_modification}c. At the pixel corner (center of the plot) the electric field lines strongly point vertical in this configuration with a potential minimum just under the deep p-well in the pixel corner. In the pixel corner the lateral field is strongly reduced which delays the drift towards electrodes and accumulates electrons under the deep p-well. After irradiation this charge gets trapped and the overall signals induced on the adjacent electrodes is reduced. Figure~\ref{fig:tcad-field}b shows the potential and field lines when a 4$\mu$m wide extra deep p-well surrounds the pixel edge (figure \ref{fig:process_modification}e). The extra p-well generates a stronger lateral field in the pixel corner, which enhances the drift of ionisation charge towards electrodes. Further down in the bulk of the sensor there is no appreciable difference in potential or field lines anymore on Czochralski substrates.

\begin{figure}[!htb]
 \centering
    \includegraphics[width=.8\textwidth]{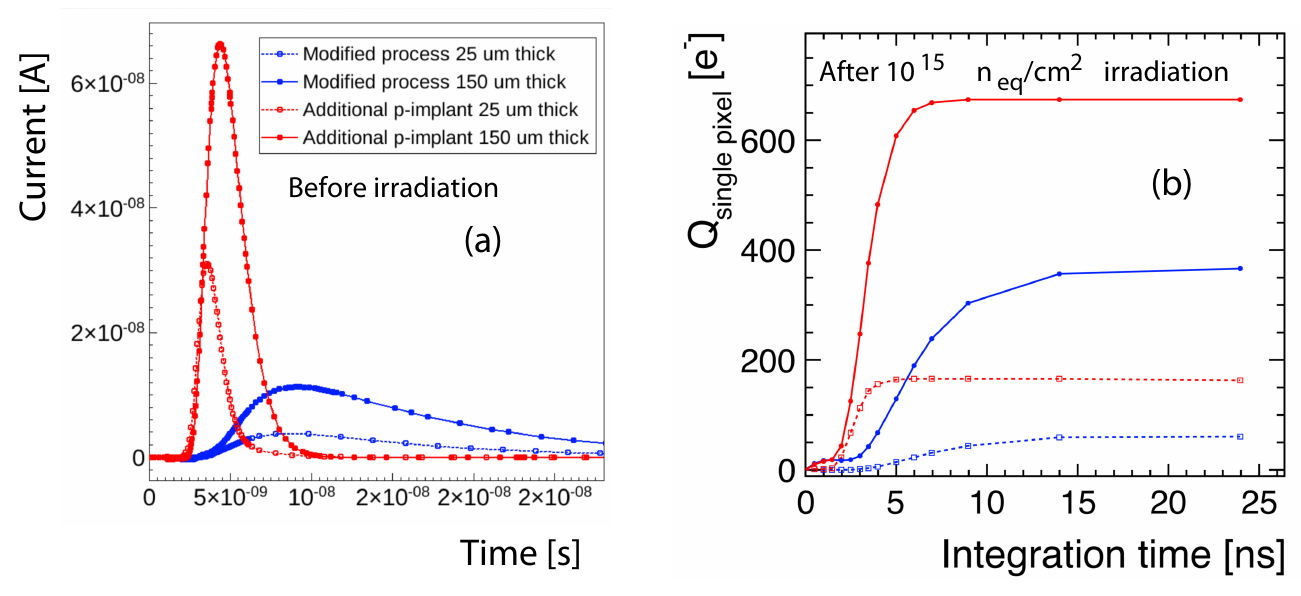}
    \caption{(a) Transient current signal before irradiation for sensors with continuous n$^-$ layer (blue curve) and sensors with extra deep-p-well (red curve). (b) Integrated current on a single pixel in the same cases after irradiation to $1\times10^{15}$ 1 MeV n$_{eq}$/cm$^{2}$ fluence.}
    \label{fig:tcad-signal}
\end{figure}

\begin{figure}[]
 \centering
    \includegraphics[width=.95\textwidth]{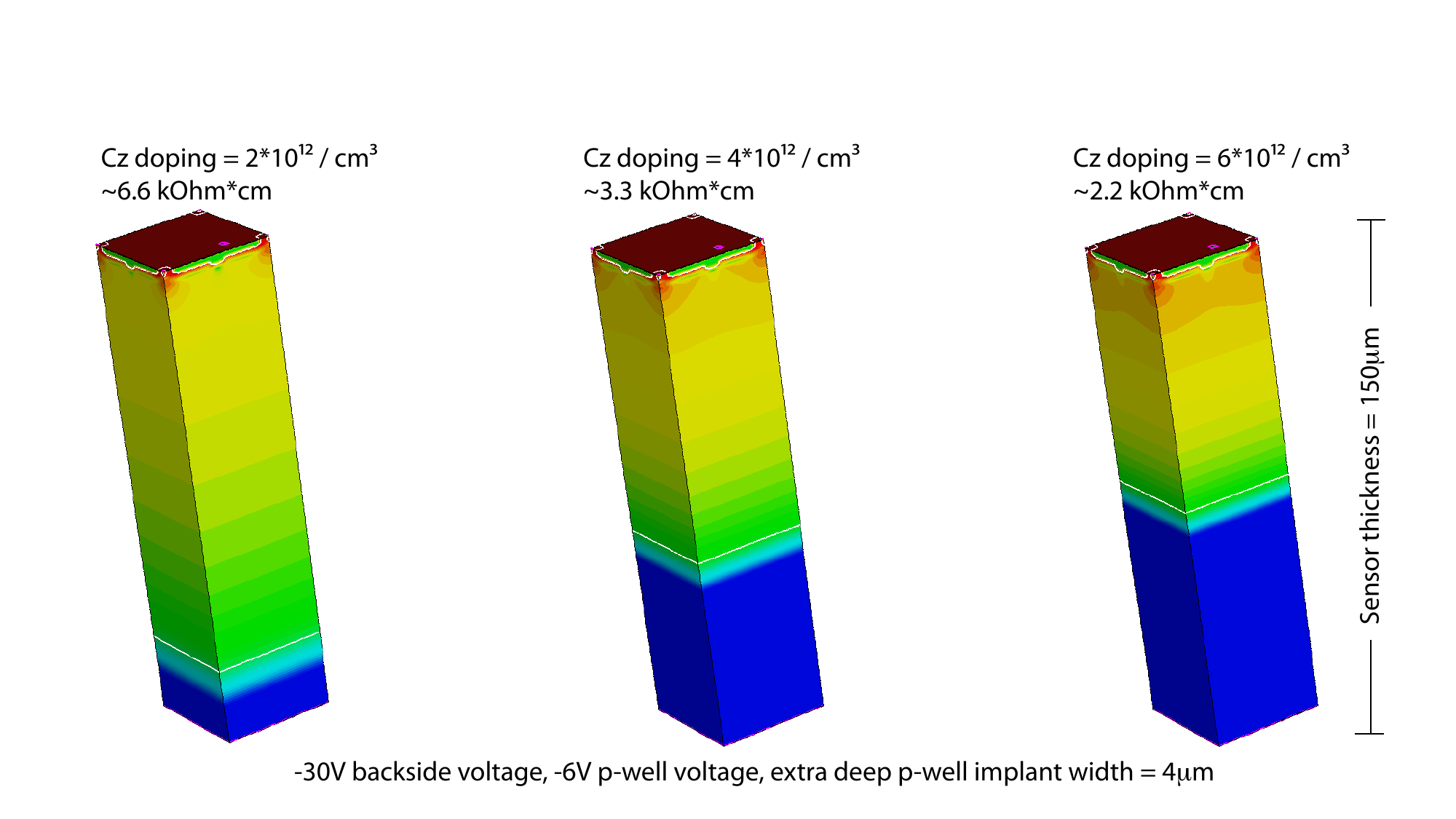}
    \caption{Electric field as function of Cz-substrate depth for three assumed boron doping levels of the high-resistivity substrate.}
    \label{fig:depletion-Cz}
\end{figure}

To study the signal amplitude and timing of different sensor implant and substrate configurations we simulate transients of minimum ionising charge particles transversing the detector in the pixel corner. This represents the most difficult case for detection efficiency as the signal is shared between four pixels and ionisation charge is deposited in the sensor volume with slowest charge collection. To model the effect of radiation damage on the signal response, in particular the signal charge, defect levels are introduced in TCAD simulation as in reference \cite{moscatelli2016}. Figure \ref{fig:tcad-signal}a shows the transient current for different implant configurations with the charge deposition close to the pixel corner, i.e. furthest away from the electrode. Blue curves illustrate the signal response for a sensor configuration with continuous  n$^-$ layer whereas red curves show the response of sensors with the additional 4$\mu$m wide extra deep-p-well along the sensor edge. A substrate bias voltage of -50~V was assumed in both cases, resulting in full depletion of a  25~$\mu$m thick epitaxial layer but under-depletion of the Czochralski sensor (see center plot of figure ~\ref{fig:depletion-Cz}). The additional deep p-well around the pixel edges significantly increases the current amplitude and results in a faster charge collection on both sensors. In addition to the faster signal, the thick sensor also provides an increase in current amplitude by a factor two. In figure \ref{fig:tcad-signal}a it should be noted that the induced current signal only starts approximately 2~ns after the simulated particle transient. Simulations have shown that this is due to the necessary drift time of the first ionization clusters from the pixel corner under the deep p-well to reach the electrode region. If this transient particle crosses in a distance of up to $\approx 8~\mu$m of the electrode the current is induced immediately.

The benefit of thick Czochralski substrate becomes evident in figure \ref{fig:tcad-signal}b, which shows the integrated single pixel charge for the different sensor configurations after irradiation to $1\times10^{15}$ n$_{eq}$/cm$^{2}$ fluence. Thick Czochralski-type sensors using an extra deep p-well around the pixel edges provide a factor of almost 10 larger single pixel charge than the original pixel design with continuous n$^-$ layer on 25$\mu$m thick epitaxial silicon. The signal response of sensors using a gap in the n$^-$ layer (e.g. figure \ref{fig:process_modification}a or \ref{fig:process_modification}d) is nearly identical to sensors with extra deep p-well in simulation results, so only one example is shown. Earlier measurements on prototype sensors \cite{dyndal2019} also show very similar results for sensors with a gap in the n$^-$ layer and sensors with the additional deep p-well implant after irradiation.

The actual depletion layer thickness will depend primarily on two factors: the resistivity of the Cz-substrate or epitaxial layer and the applied sensor substrate bias. The latter is limited by punch-through current between deep p-well and p-type substrate, which are separated by the n$^-$ layer, as they are operated with significant voltage difference. Figure~\ref{fig:depletion-Cz} shows the electric field strength as function of substrate depth for three different assumed bulk boron-doping levels of $2\times10^{12}$/cm$^{3}$, $4\times10^{12}$/cm$^{3}$, $6\times10^{12}$/cm$^{3}$, which translates to resistivity of 6.6~k$\Omega$cm, 3.3~k$\Omega$cm and 2.2~k$\Omega$cm respectively. The vertical axis ($z$-axis) corresponds to the simulated detector thickness of 150~$\mu$m. The substrate bias voltage was fixed to -30~V in the simulation while the p-well voltage is fixed at -6~V. The white line marks the limit of depletion zone. The depletion ranges from nearly full depletion at the highest resistivity to approximately half depletion at the lowest resistivity. 

\begin{figure}[]
 \centering
    \includegraphics[width=.95\textwidth]{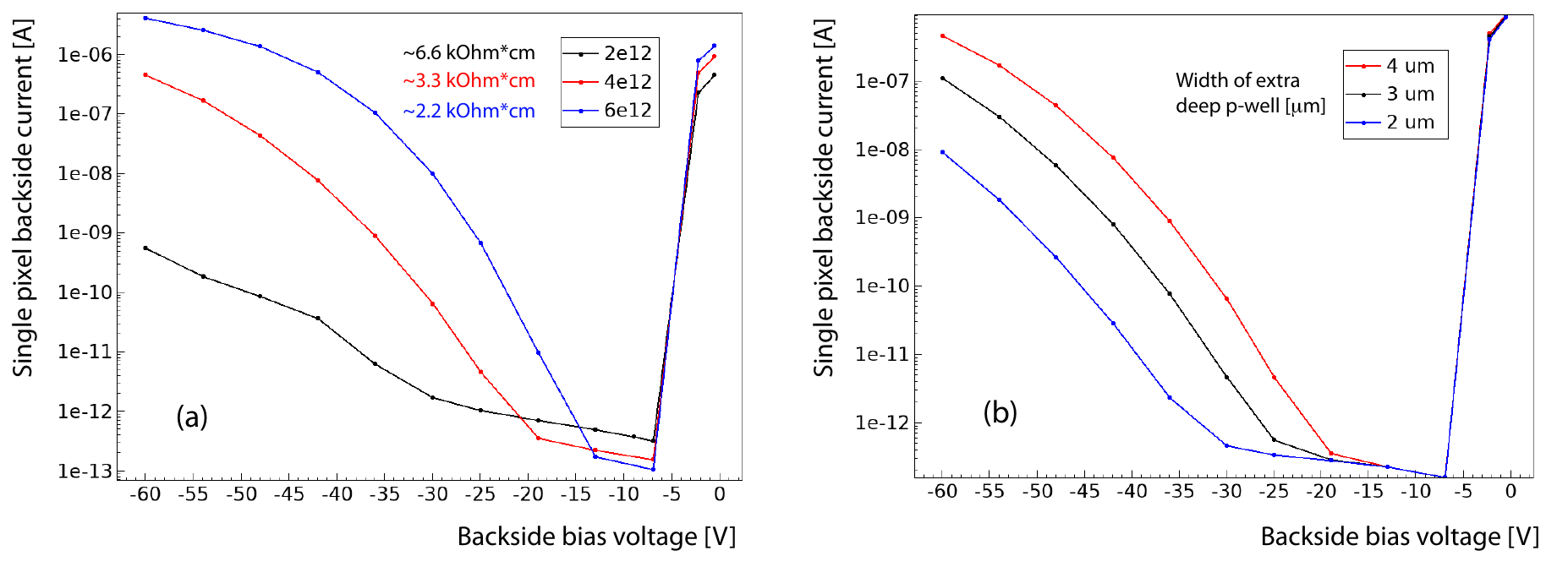}
    \caption{Punch-through current as function of substrate voltage for different substrate resistivities at 4$\mu$m additional deep p-well implant width (a) and for different additional deep p-well implant widths at a substrate resistivity of 3.3~k$\Omega$cm (b).}
    \label{fig:punch-thru}
\end{figure}

The punch-through current between the deep p-well and substrate as a function of substrate voltage was simulated in TCAD for different implant geometries and substate resistivities. We find no significant punch-through current for implant configurations with a continuous  n$^-$ layer (figure~\ref{fig:process_modification}c) up to voltages of approximately -50~V. For configurations with either a gap in the n$^-$ layer (figure~\ref{fig:process_modification}d) or an additional deep p-well (figure~\ref{fig:process_modification}e) along the pixel edge, punch-through occurs as lower substrate bias, hence effectively limiting the depletion zone by limiting the operational voltage. Figure~\ref{fig:punch-thru} shows the pixel substrate current as a function of substrate bias. The p-well voltage is fixed at -6~V.  Figure~\ref{fig:punch-thru}a shows the current for an implant configuration with 4$\mu$m wide additional deep p-well (figure~\ref{fig:process_modification}e) on a substrate with three different resistivities of 6.6~k$\Omega$cm, 3.3~k$\Omega$cm and 2.2~k$\Omega$cm respectively. On lower resistivity materials the substrate current significantly increases beyond -15~V to -20~V while the highest resistivity material provides the largest operational range for the substrate voltage. The current increase at low voltage is due to the p-well  being biased at -6~V. A spatial analysis of the hole current in the substrate bulk shows that punch-through occurs primarily at the pixel edge in the region of the additional deep p-well where the deep-p-well and substrate are minimally separated by the n$^-$ layer. For this reason we also expect a dependence of punch through on different additional deep p-well implant widths, which is shown in figure~\ref{fig:punch-thru}b for a substrate resistivity of 3.3~k$\Omega$cm. As expected the punch-through voltage increases when the additional deep p-well width (or gap width in n$^-$ layer) is reduced. A reduced width however also reduces the lateral electrical field strength at the pixel edges, which affects charge collection efficiency after irradiation. The optimal width is a trade-off between maximum depletion thickness through highest substrate bias, charge collection in the pixel corners and manufacturing limitations for different implants.

\section{Measurements on unirradiated and irradiated MALTA-Cz sensors}

In the following measurements we will analyse the performance of MALTA sensors produced on epitaxial and Czochralski substrates before and after neutron irradiation. Sensors have been irradiated with neutrons at the Triga reactor at the Institute Jo\v{z}ef Stefan, Ljubljana, Slovenia \cite{Snoj:2012dib,Ambro2017}. After irradiation the sensors have been kept cold except for mounting to PCB and wire bonding. After mounting to PCB the sensors were tested in laboratory tests and beam tests. During the initial tests we have verified the operation of sensors as function of substrate voltage. During the tests the sensors are configured and readout, p-wells are biased at -6V and analog and digital V$_{dd}=$1.8V are supplied.  

\subsection{IV characteristics of irradiated MALTA-Cz sensors}

\begin{figure}[!htb]
 \centering
    \includegraphics[width=.65\textwidth]{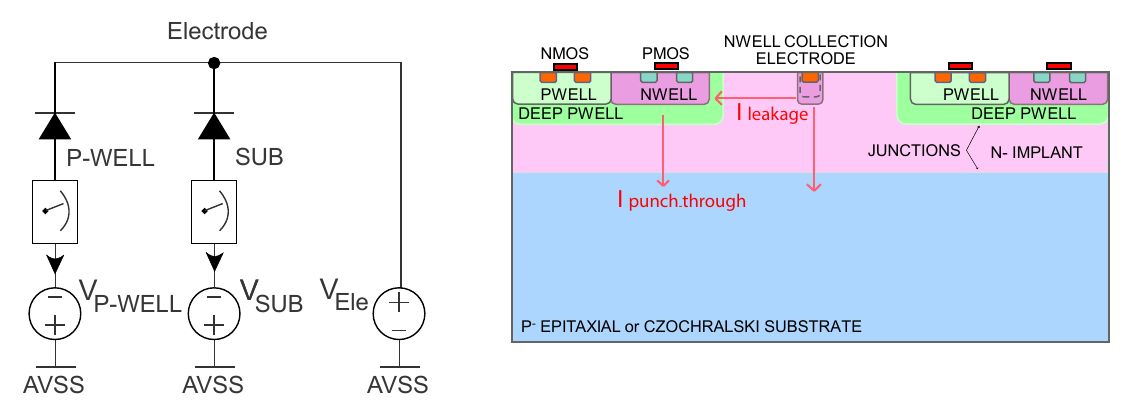}
    \caption{Measurement setup for current-voltage characteristics of substrate and p-well currents.}
    \label{fig:IVmeas}
\end{figure}

Given the importance of electrical separation between p-well and substrate through the n$^{-}$ layer we have tested the  IV characteristics on unirradiated and  irradiated sensors at different temperatures. For this measurement the sensors current was measured on the substrate contact and the p-well contact as a function of substrate bias voltage. The p-well voltage was kept fixed at -6~V, the n$^{+}$ collection electrode was biased at +0.8~V. The substrate and p-well currents include the full active matrix of 18.3$\times$18.3~cm$^{2}$ as well as the surrounding p-well ring currents close to the scribe line of the sensor. Due to the biasing and reset circuit connected to the n$^{+}$ collection electrode it is not possible to measure the current at the n$^{+}$ electrode. Figure~\ref{fig:IVmeas} shows the sensor biasing schematics during the IV measurements, where the substrate voltage is increased starting with the p-well voltage of -6V. The figure also denotes the punch-through current flowing between p-well and p-type substrate across the n$^{-}$ layer once the potential barrier is overcome when a sufficiently large potential difference between p-well and substrate is reached. The leakage currents flowing across the junctions between n$^{+}$ electrode and p-well, and also n$^{-}$ layer and p-substrate, contribute to p-well and substrate current. The substrate current is the sum of punch-through current and league current across the n$^{-}$ layer/p-substrate junction. The p-well current as shown in figure~\ref{fig:IVmeas} is determined by the leakage current across n$^{+}$ electrode/p-well junction minus the punch-through current. We estimate the total leakage current as $I_{leak} = I_{sub}+I_{p-well}$ and the punch-through current as $I_{pt} \approx \frac{1}{2} (I_{sub}-I_{p-well})$. The approximation for the punch-through current holds true as long as the difference in leakage currents is small, i.e. without break down.

\begin{figure}[!htb]
 \centering
    \includegraphics[width=.85\textwidth]{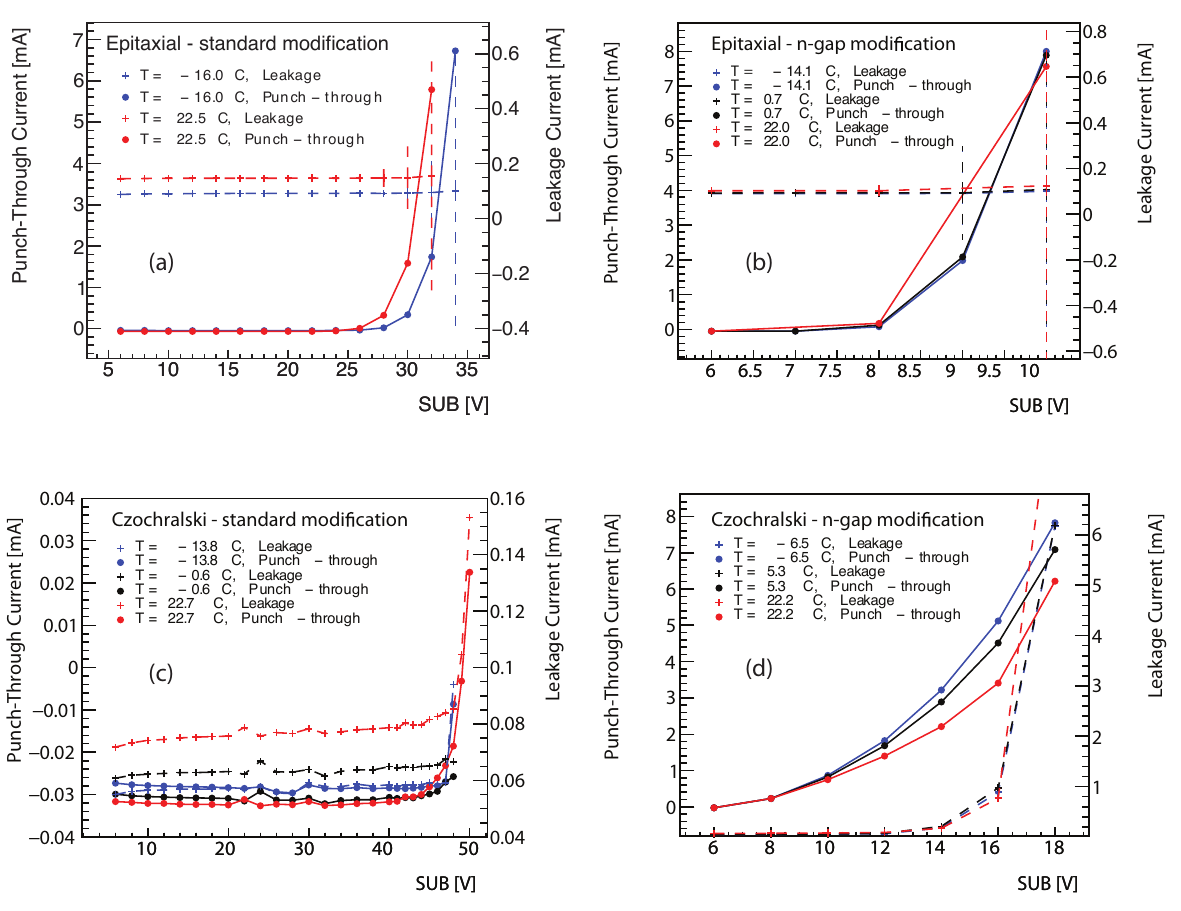}
    \caption{Punch-through and leakage current as a function of substrate voltage with respect to ground with p-well at -6V on unirradiated sensors. Figures (a) and (b) show the IV characteristics of epitaxial sensors, figures (c) and (d) of Czochralski sensors. Sensors shown in figures (a) and (c) have a continuous n$^-$ layer and figures (b) and (d) show the IV characteristics of sensors with n$^-$ gap. IV curves are given for different temperatures.}
    \label{fig:IV-unirrad}
\end{figure}

Figure~\ref{fig:IV-unirrad} shows the punch-through (left axis) and leakage current (right axis) as function of substrate voltage ($|V_{sub}|$) with respect to ground for unirradiated epitaxial and Czochralski sensors. Figures (a) and (c) show the IV characteristics of epitaxial and Czochralski sensors with continuous n$^-$ layer.  Figures (b) and (d) show the IV characteristics of epitaxial and Czochralski sensors with n$^-$ gap. Sensors with continuous n$^-$ layer can be operated up to -30~V on epitaxial and up to -47~V on Czochralski substrates.  The introduction of a 4$\mu$m gap along the pixel edges in the n$^-$ layer significantly lowers this operational limit: on epitaxial substrates we observe significant punch-through current within a few volts of potential difference between p-well and substrate, on Czochralski substrates we can operate the sensor up to a potential difference between p-well and substrate of $\approx$10~V. 
The reduction of punch-through voltage is due to the significant decrease in potential barrier between substrate and p-well caused by the gap in the n-layer. As figure~\ref{fig:punch-thru} shows, the geometry of this gap, the choice of substrate and the implantation profile depths for deep p-well and n$^-$ layer influence the onset of punch-through. 

\begin{figure}[!htb]
 \centering
    \includegraphics[width=1.05\textwidth]{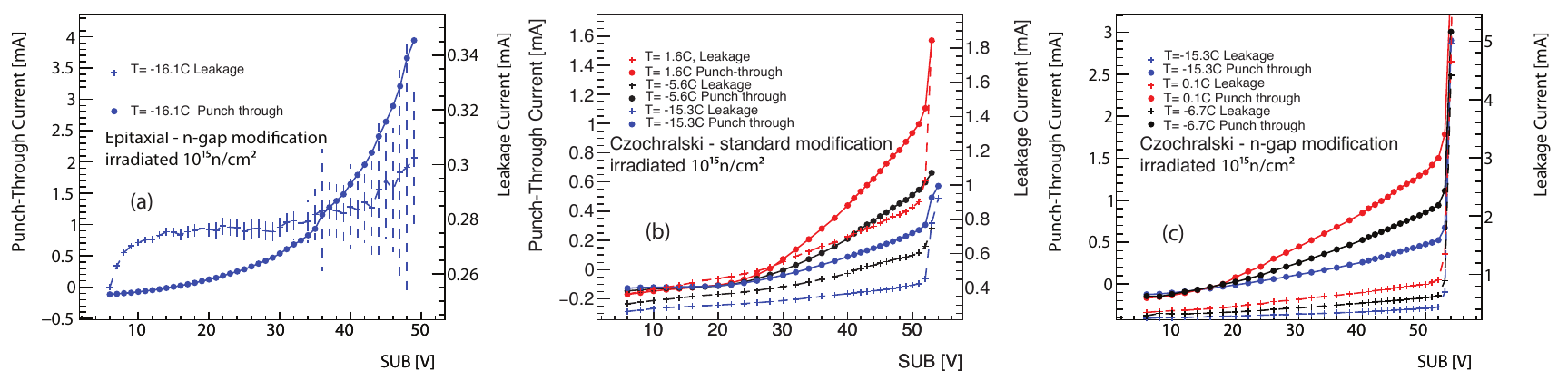}
    \caption{Punch-through and leakage current as a function of substrate voltage with respect to ground with p-well at -6V on $1\times10^{15}$n$_{eq}$/cm$^{2}$ neutron irradiated sensors. Figure (a) shows the IV characteristics of epitaxial sensors, figures (b) and (c) of Czochralski sensors. Sensors shown in figures (a) and (c) have a n$^-$ gap and sensors shown in figure (b) have a continuous n$^-$ layer. IV curves are given for different temperatures.}
    \label{fig:IV-1e15-pt}
\end{figure}

Figure~\ref{fig:IV-1e15-pt} shows the IV characteristics after $1\times10^{15}$ n$_{eq}$/cm$^{2}$ neutron irradiation at different temperatures for epitaxial sensors with n$^-$ gap (a), for Czochralski sensors with continuous n$^-$ layer (b) and for Czochralski sensors  with n$^-$ gap (c). The temperature dependence of the IV curves indicate a clear contribution to the currents from bulk-damage generation current as expected after neutron irradiation. Figures (a) and (c) show a significant increase in punch-through voltage after irradiation for sensors with n$^-$ gap. The break-down occurs on sensors with n$^-$ gap after irradiation at potential differences between p-well and substrate that match sensors with a continuous n$^-$ layer, hence operation up to $\approx$ -50~V is possible also on sensors with n$^-$ gap. Similar behaviour is observed for sensors with extra deep p-well. The model of a ``double junction'' as described in section~\ref{process} can offer a possible explanation: irradiation causes the creation of deep-level donors, which adds effective n-doping in the gap near the p-well. This further increases the potential barrier between p-well and p-type substrate. This space charge acts like extra n-doping which ``narrows'' the gap after irradiation.

\subsection{Sensor efficiency before and after irradiation} \label{sec:eff}

To measure the sensor detection efficiency for charged particles before and after irradiation, the MALTA sensors are arranged in a 4-layer telescope and the detection efficiency of one of the sensors (``DUT'') is measured while the other sensors are used to reconstruct the particle trajectory. The data presented in this study were recorded at the DESY test beam facility at DESY Hamburg, Germany using a 4~GeV electron beam. Three 100~$\mu$m thick MALTA epitaxial sensors serve as reference planes. Electrons pass first through one MALTA reference plane before they pass through the DUT at 18~mm distance. Downstream of the DUT reference planes are located at distances of 2~mm and 19~mm from the DUT respectively. Unirradiated DUTs are operated at room temperature, while irradiated DUTs are contained in a cold box which encloses the cooled DUT sensor (-14$^{\circ}$C).

To estimate the particle hit position on the DUT, the trajectory is reconstructed from the three reference planes and then interpolated on the DUT plane. The hit prediction accuracy is limited by multiple scattering due to the low energy electron beam. The track trajectory calculation uses the material description of the DUT and all telescope planes as well as the electron beam energy to estimate multiple scattering when applying the General Broken Lines (GBL) formalism~\cite{kiehn_moritz_2019_2586736,Kleinwort:2012np}. The hit detection efficiency is defined as the fraction of clusters on the DUT matched to telescope tracks over the total number of tracks. The DUT hit is matched to a track if the distance between the track interpolation position and the center position of the cluster is smaller than 60~$\mu$m.

The detection efficiency of irradiated epitaxial sensors using the MALTA front-end design has already been reported in reference \cite{dyndal2019}. This report focuses on the performance obtained with the full-size MALTA sensor produced on Czochralski substrate applying the same implant geometries as reported in \cite{dyndal2019} but on a full-size matrix. In comparison to reference \cite{dyndal2019} it is essential to note that the full-size MALTA sensor of this report uses the so-called "standard" front-end design with minimal size transistors featuring a lower gain and therefore limitations to the lowest achievable threshold. Irradiated sensors ($1\times10^{15}$ n$_{eq}$/cm$^{2}$) produced on epitaxial substrate showed significant inefficiency in the corner region even with n$^-$ gap or extra deep p-well along the pixel edge due to the low-gain front-end and its minimal threshold limitations.

\begin{figure}
    \centering
    \includegraphics[width=.8\textwidth]{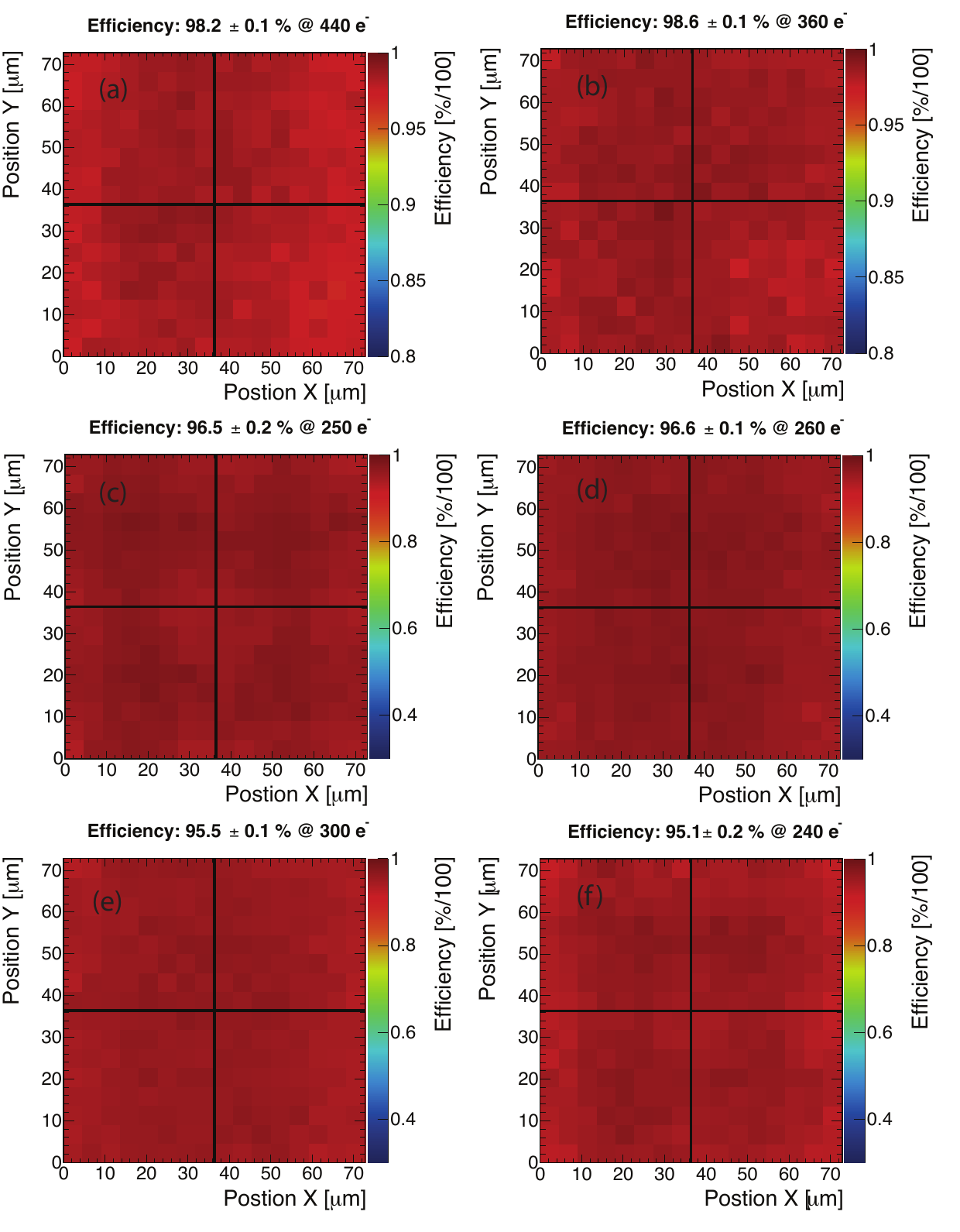}
    \caption{In-pixel 2D efficiency maps shown as a stacked $2\times2$ pixel group for (a) unirradiated epitaxial sensor with n$^-$ gap, (b) unirradiated Czochralski sensor with n$^-$ gap, (c) $1\times10^{15}$ n$_{eq}$/cm$^{2}$ irradiated Czochralski sensor with continuous n$^-$ layer,  (d) $1\times10^{15}$ n$_{eq}$/cm$^{2}$ irradiated Czochralski sensor with n$^-$ gap, (e) $1\times10^{15}$ n$_{eq}$/cm$^{2}$ irradiated Czochralski sensor with extra deep p-well, (f) $2\times10^{15}$ n$_{eq}$/cm$^{2}$ irradiated Czochralski sensor with n$^-$ gap.}
    \label{fig:eff2x2}
\end{figure}

Figure~\ref{fig:eff2x2} shows that the efficiency in the corner can be fully recovered when moving to Czochralski substrate. The figure shows the in-pixel efficiency distribution through stacking of all sector pixels in a 2$\times$2-pixel plot. The beam test telescope is used to predict the track position, shown as horizontal and vertical axis. The location where all 4 adjacent pixel corners meet is in the center of the plot, any efficiency loss in the pixel corners can be identified as drop of efficiency in the plot center. The overall efficiency together with its statistical error is given in the plot title together with the threshold at data taking. Unirradiated sensors (figures~\ref{fig:eff2x2} (a) and (b)) are operated at $V_{sub} = -6$~V, $1\times10^{15}$ n$_{eq}$/cm$^{2}$ irradiated sensors (c-e) are operated at $V_{sub} = -50$~V, and a $2\times10^{15}$ n$_{eq}$/cm$^{2}$ irradiated sensor (f) is operated at $V_{sub} = -55$~V. All irradiated sensors are operated cold at  $\approx -14^{\circ}$C, the efficiency was measured on sector 2 of all MALTA sensors. The Czochralski sensors are thinned to 300$\mu$m thickness. Unirradiated sensors register minimum ionising particles with $>98$\% efficiency, $1\times10^{15}$ n$_{eq}$/cm$^{2}$ irradiated Czochralski sensor with $95.5$\% to $96.6$\% efficiency, the $2\times10^{15}$ n$_{eq}$/cm$^{2}$ irradiated Czochralski sensor achieves an efficiency of $95.1$\%. The plots also indicate that there is no substantial efficiency drop in the pixel corner region after irradiation with Czochralski sensors due to their higher signal amplitude.

\begin{figure}
    \centering
    \includegraphics[width=.9\textwidth]{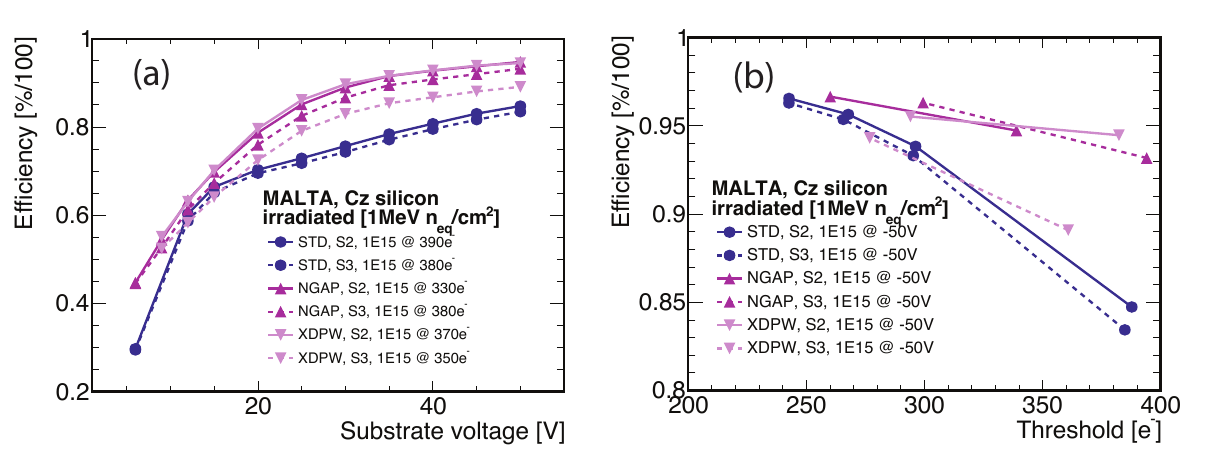}
    \caption{Sensor efficiency at elevated thresholds for $1\times10^{15}$ n$_{eq}$/cm$^{2}$ irradiated Czochralski sensor with continuous n$^-$ layer,  irradiated Czochralski sensor with n$^-$ gap and irradiated Czochralski sensor with extra deep p-well. Plot (a) shows the efficiency as a function of substrate voltage with constant threshold across the samples, plot (b) show the efficiency as a function of threshold with constant substrate voltage of -50~V.}
    \label{fig:eff-type}
\end{figure}

To investigate the effect of different pixel implant designs (cf. figures~\ref{fig:process_modification} c-e) on overall performance, we measure the efficiency at medium thresholds levels. The threshold ($\approx$ 350 e$^{-}$) was chosen such that hits in the pixel corners exhibit visible inefficiency. The threshold values are adjusted to be as uniform as possible across the different sensors. By comparing the overall measured efficiency at this medium threshold levels we can investigate the effectiveness of the different implant designs for charge collection in the pixel corners after irradiation. Figure~\ref{fig:eff-type} shows sensor efficiencies for $1\times10^{15}$n$_{eq}$/cm$^{2}$  irradiated Czochralski sensors with continuous n$^-$ layer (marked as ``STD''),  sensors with n$^-$ gap (marked as ``NGAP'') and sensors with extra deep p-well (marked with ``XDPW''). ``S2'' denotes data taken in MALTA sector 2 with the maximum extent of deep p-well, ``S3'' denotes data taken in the sector with a reduced deep p-well. Figure~\ref{fig:eff-type}a shows the efficiency as a function of substrate voltage with constant threshold, which shows that corner efficiency is better on sensors with either a gap in the n$^-$ layer or an extra deep p-well along the pixel edge. This behaviour has been predicted in TCAD simulation qualitatively as shown in figure~\ref{fig:tcad-signal}. Figure~\ref{fig:eff-type}b shows that sensors with this additional modification also provide a wider operational range for threshold settings while maintaining a higher efficiency than Czochralski sensors with a continuous n$^-$ layer.  The improvements in corner efficiency, which has previously been shown on epitaxial sensors, also apply qualitatively to sensors produced on Czochralski substrate.

\begin{figure}
    \centering
    \includegraphics[width=.6\textwidth]{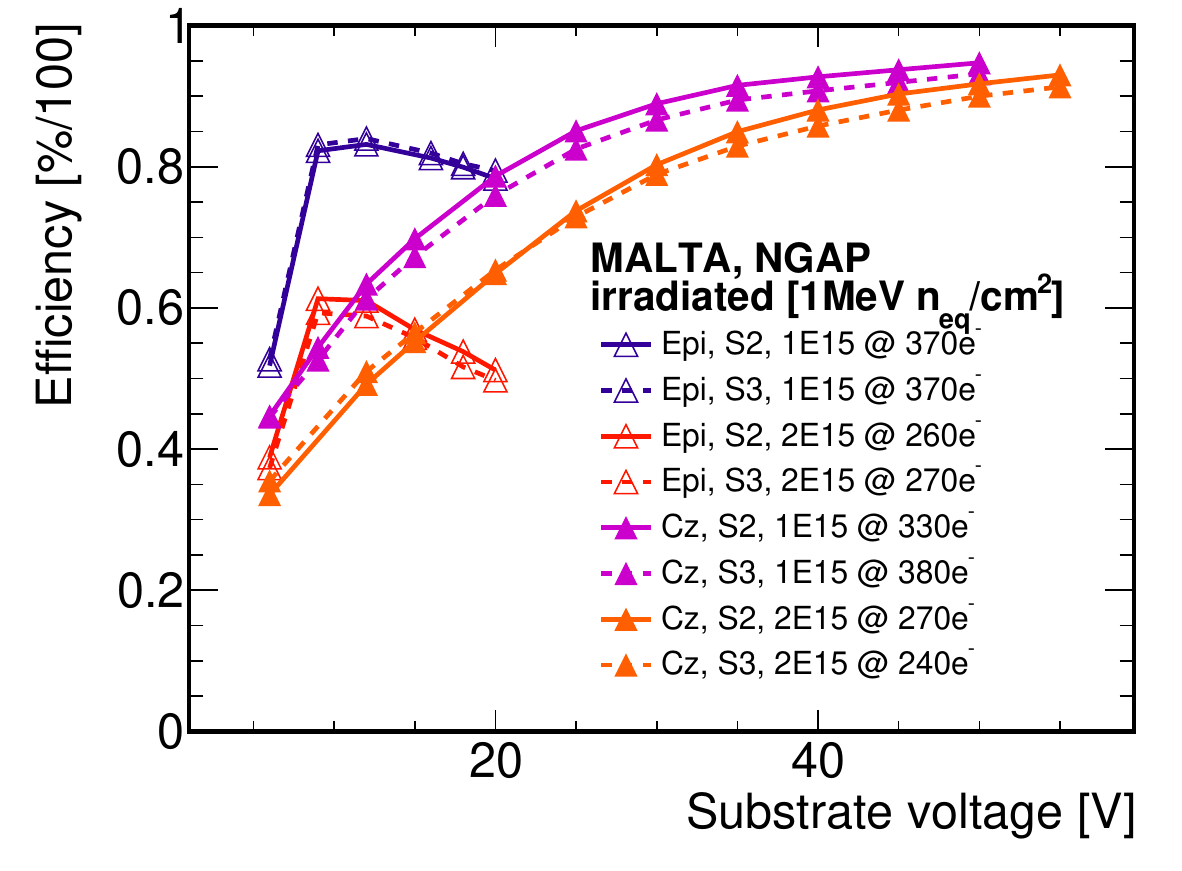}
    \caption{Sensor efficiency for $1\times10^{15}$ n$_{eq}$/cm$^{2}$ and $2\times10^{15}$ n$_{eq}$/cm$^{2}$  irradiated epitaxial and Czochralski sensors with n$^-$ gap as a function of threshold.}
    \label{fig:eff-epivscz}
\end{figure}

 Figure~\ref{fig:eff-epivscz} illustrates the different behaviour of epitaxial and Czochralski sensors as a function of substrate bias after irradiation to $1\times10^{15}$ n$_{eq}$/cm$^{2}$ and $2\times10^{15}$ n$_{eq}$/cm$^{2}$. For this comparison all sensors are operated with same front-end parameters, to minimise differences from front-end amplifier response. While these settings produce low thresholds, they are not optimised for maximum efficiency on each sample. The epitaxial sensors achieve a maximum efficiency at $\approx -12$~V, and at higher voltages the efficiency decreases. The efficiency decrease can be understood through electric field strength simulation at the pixel boundaries. With increasing bias voltage the vertical field strength further increases however the later field decreases which reduces the effectiveness of charge collection near the boundary.
 
Contrary to epitaxial sensors, the efficiency on Czochralski sensors increases substantially with substrate voltage as the depleted zone in the high resistivity substrate increases and with it the ionisation signal. While we have comparable settings for the front-end amplifiers for the circuit, with the exception of the capacitive load on the amplifier input, the efficiency is significantly better for the Czochralski sensors than for the epitaxial sensors due to the larger signal charge. This is particularly evident for $2\times10^{15}$ n$_{eq}$/cm$^{2}$ irradiated sensors.

\subsection{Measurements of cluster size before and after irradiation}

The size of the cluster, i.e. the number of adjacent pixels firing after a charged particle transverses the detector, is important for several applications. In applications where the hit position reconstruction is improved by using centre-of-gravity calculations, a wider cluster with charge sharing between pixels is desirable. In applications with very high hit rates and/or high radiation environments typically narrow clusters are preferred to improve two-track separation and increase the individual pixel signal-to-noise ratio. We have measured the dependence of cluster size on substrate bias for different implant configurations, on epitaxial and Czochralski substrates, and before and after irradiations. The measurements give an indication of how the implant design and substrate choice influences charge sharing, and hence can be used to optimise the sensor for different applications.

\begin{figure}
    \centering
    \includegraphics[width=.8\textwidth]{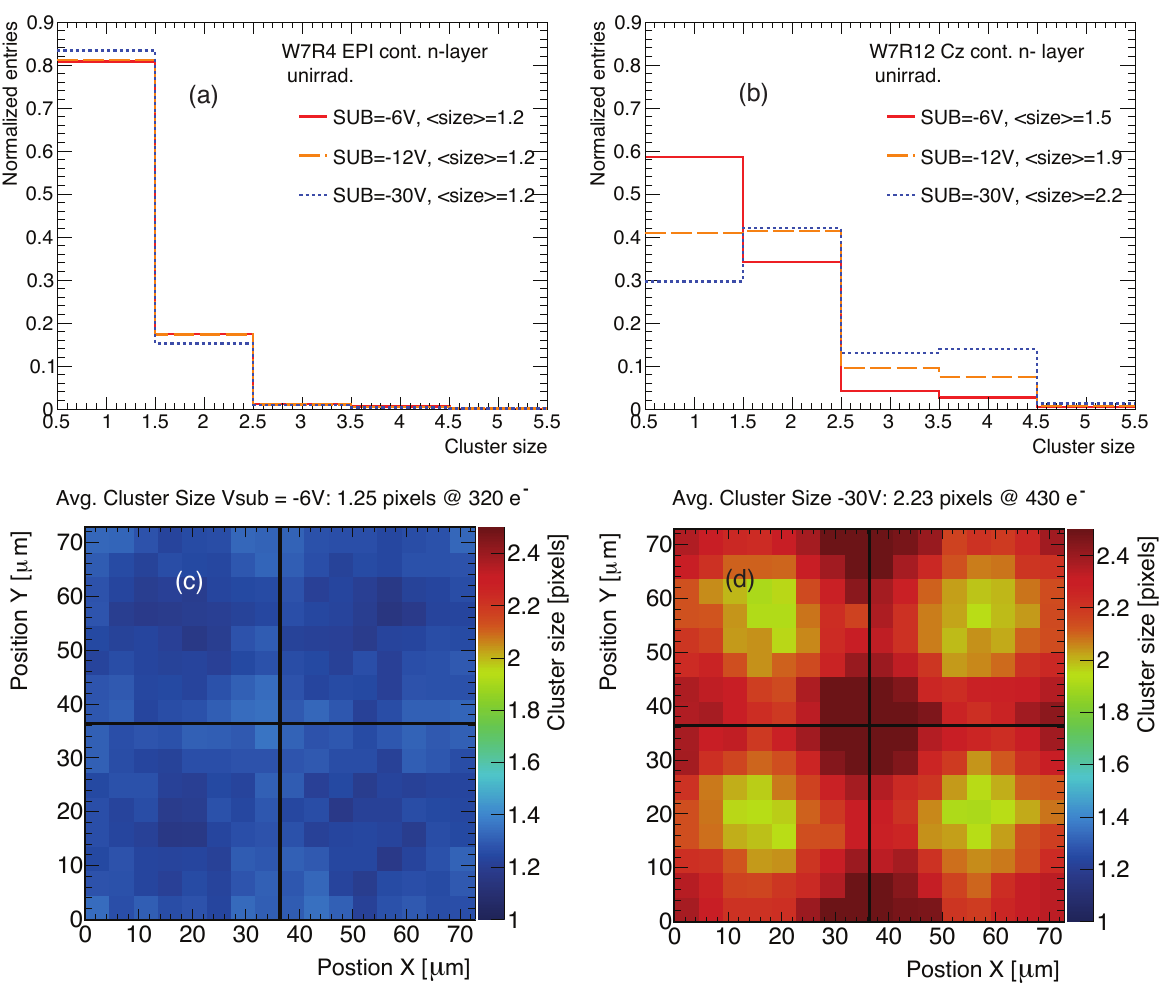}
    \caption{Cluster size of unirradiated epitaxial (plots (a) and (c)) and  Czochralski (plots (b) and (d)) sensors with continuous n$^-$ layer. Plots (a) and (b) shows the number of hit pixels in a cluster as function of substrate bias voltage. Plot (c) and (d) shows the number of hit pixels in a cluster as function of X/Y track position in a 2$\times$2 pixel array. The epitaxial sensor is operated at a threshold of 320 e$^{-}$ and a substrate bias of -6~V, the Czochralski sensor is operated at a threshold of 430 e$^{-}$ and a substrate bias of -30~V}
    \label{fig:Clustersize}
\end{figure}

Figure~\ref{fig:Clustersize} shows the cluster size measured in beam tests on MALTA sensors produced on epitaxial (figures ~\ref{fig:Clustersize}a and ~\ref{fig:Clustersize}c) and Czochralski (figures ~\ref{fig:Clustersize}b and ~\ref{fig:Clustersize}d) substrates with a continuous n$^-$ layer. While the epitaxial sensor shows a nearly constant cluster size with increasing substrate voltage, the cluster size on the Czochralski increases significantly with increasing substrate bias. As the substrate bias increases, the depleted thickness of the substrate increases and the induced signal spreads over more pixels also due to diffusion. In particular for hits in the pixel corners, the signal is spread over two, three or four adjacent pixels. This is nicely illustrated in figure~\ref{fig:Clustersize}d, where the cluster size substantially increases in regions near the pixel edge and in particular in the corners. The center of the pixel registers the smallest cluster size while the pixel corner region (center of plot) has the highest cluster size. In comparison, on epitaxial sensors (figure~\ref{fig:Clustersize}c) the average cluster size is smaller overall and rather uniform as a function of hit position. Especially for perpendicular tracks charge sharing improves spatial resolution through hit interpolation. Therefore sensors on Czochralski substrate have a clear advantage over epitaxial substrate sensors with the identical implant design.

\begin{figure}
    \centering
    \includegraphics[width=.8\textwidth]{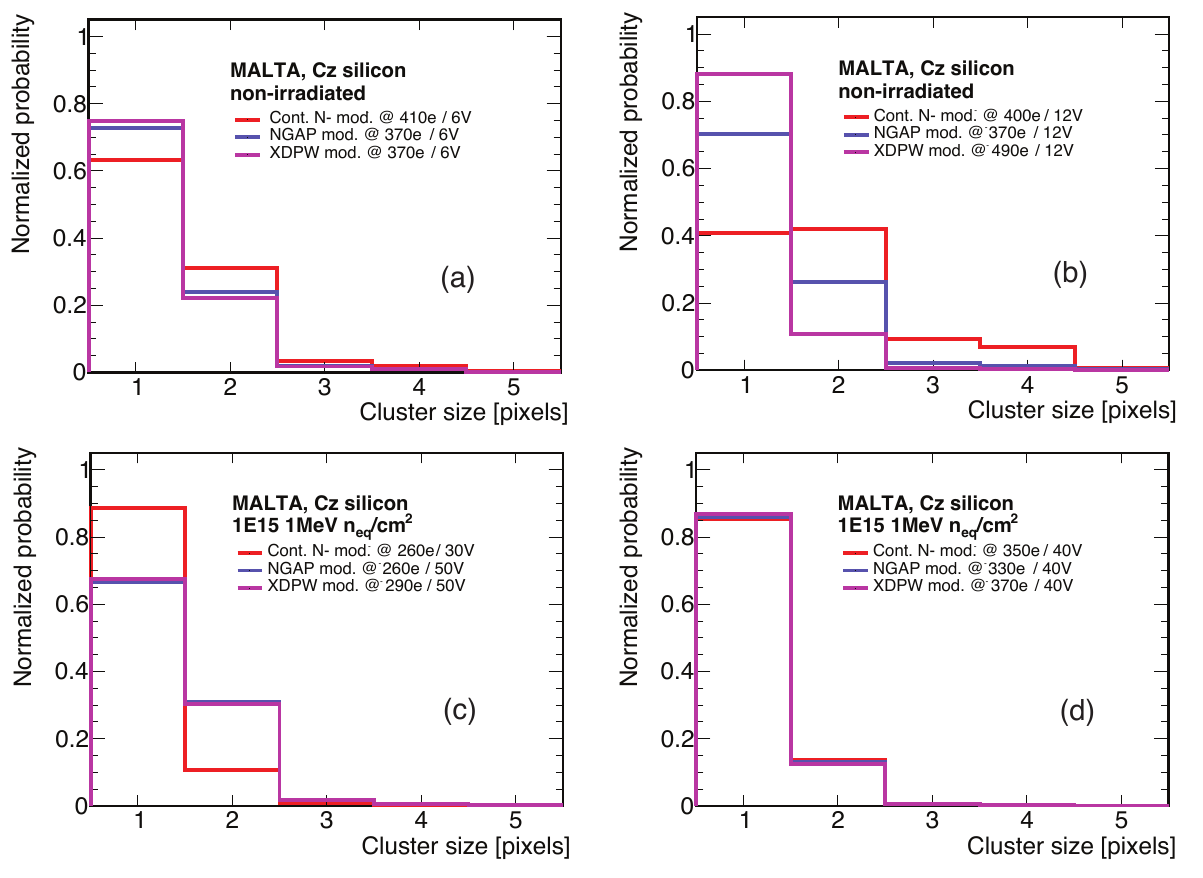}
    \caption{Cluster size of unirradiated (top row) and $1\times10^{15}$ n$_{eq}$/cm$^{2}$ irradiated (bottom row) Czochralski sensors with continuous n$^-$ layer, with n$^-$ gap and with extra deep p-well implant designs. The normalised cluster width distributions are recorded at different thresholds and substrate bias voltages as given in the figure legends.}
    \label{fig:cluster-irrad}
\end{figure}

Figure~\ref{fig:cluster-irrad} illustrates the influence of different implant designs, a continuous n$^-$ layer, n$^-$ gap options and an extra deep p-well implant. Figures~\ref{fig:cluster-irrad}a and~\ref{fig:cluster-irrad}b show the cluster size of unirradiated Czochralski sensors at lower (a) or higher (b) substrate bias. The comparison shows that n$^-$ gap and extra deep p-well modifications to the pixel edge narrows the signal spread across pixels, which suggests that the induced signal is strongly dominated by the weighting field close to the implants (cf. figure~\ref{fig:tcad-signal}). The difference between n$^-$ gap and extra deep p-well modification in figure~\ref{fig:cluster-irrad}b is due to the difference in threshold when the data were recorded. After irradiation the situation changes significantly. At lowest threshold (figure ~\ref{fig:cluster-irrad}c) the sensor with continuous n$^-$ layer shows the narrowest cluster due to charge trapping in the corner, where only the highest signal in a cluster exceeds the threshold for hits in the pixel corner. The n$^-$ gap and extra deep p-well modifications improve charge collection from pixel corners and more pixels in a cluster exceed the threshold. At medium thresholds ($\approx$ 350 e$^{-}$) shown in figure~\ref{fig:cluster-irrad}d however the cluster size distributions of the three sensor types are nearly identical.
 
\section{Timing properties of MALTA-Cz sensors}

\begin{figure}
    \centering
    \includegraphics[width=.99\textwidth]{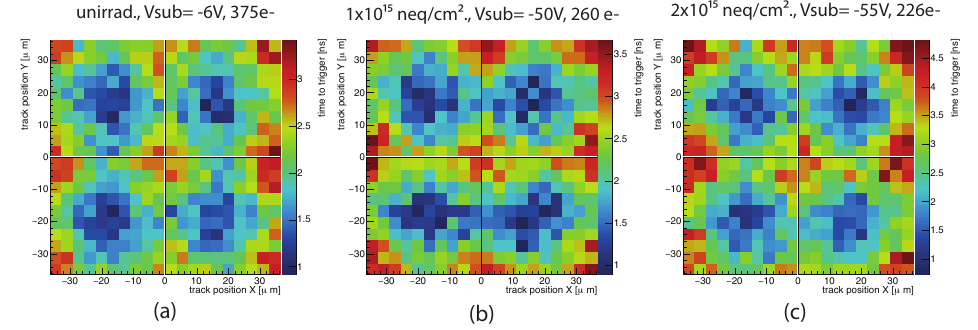}
    \caption{Signal arrival time with respect to trigger for unirradiated (a), $1\times10^{15}$ n$_{eq}$/cm$^{2}$ irradiated (b) and $2\times10^{15}$ n$_{eq}$/cm$^{2}$ irradiated (c) Czochralski sensors with n$^-$ gap modification as function of the reconstructed track position on the 2$\times$2-pixel array. The delay's constant contribution has been subtracted.}
    \label{fig:arrivaltime}
\end{figure}

In beam and source measurements the digital output signals of the MALTA sensor are recorded in FPGAs which sample the signal at 4~GHz. This allows to measure their arrival time with respect to an external trigger scintillator signal or external clock. The arrival time of the signal is determined by signal formation process (cf. figure~\ref{fig:tcad-signal}a), pre-amplifier and discriminator response and the digital signal propagation time through the double column and periphery. The signal propagation down the column to periphery was measured through test pulses to be $\approx 7$~ns for the full column height. The largest contribution to the time delay stems from time-walk in the analog front-end (maximum $\approx 2$0~ns at a threshold of 300~e$^{-}$). Figure~\ref{fig:arrivaltime} shows the arrival time of the fastest MALTA signal in a cluster with respect to a trigger scintillator (jitter $\sigma \approx$0.5~ns) as function of the predicted hit track position on the $2\times2$-pixel array. Constant delays of arrival time to trigger are subtracted, however the trigger logic adds a jitter of $\sigma=0.9$~ns when latching the asynchronous trigger signal. The average arrival time in figure~\ref{fig:arrivaltime} was measured on unirradiated and irradiated Czochralski sensors with a n$^-$ gap along the pixel edge, with substrate bias and threshold given in the figure title. The plotted time delay for hits in the pixel center is shortest, while for hits in the pixel corners the delay increases by $\approx 2$~ns on average. This delay is consistent with detector simulation and is due to the drift time of the first ionization cluster from the pixel corner to the electrode.
A strong signal is induced only when drifting charges from the pixel corner reach a distance of $\approx 5 \mu$m to $8\mu$m from the electrode. This is due to the very non-uniform weighting field in small electrode pixel designs as confirmed by field and transient current simulations.

\begin{figure}
    \centering
    \includegraphics[width=.9\textwidth]{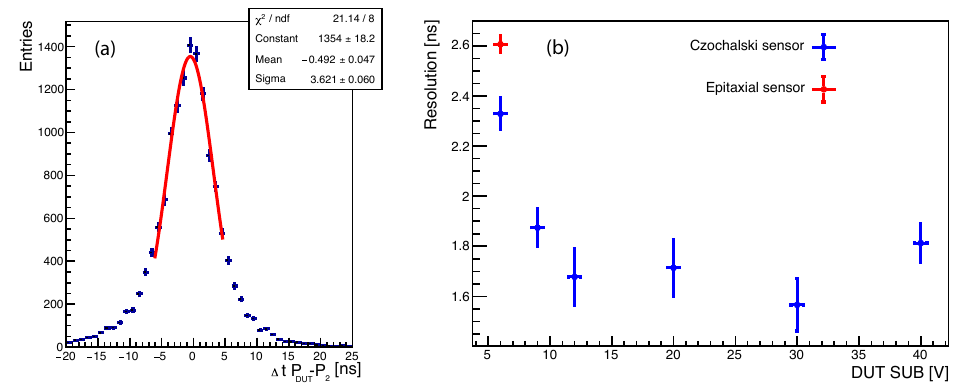}
    \caption{ (a) Time difference distribution for the fastest signals arriving from two epitaxial MALTA sensors with continuous n$^-$ layer ($t$($P_{DUT}$) minus $t$($P_{2}$)) as measured in the beam tests at DESY. Plot (b) shows the extracted time resolution of the Czochralski-MALTA sensor (blue markers) and epitaxial sensor (red marker) with continuous n$^-$ layer as function of substrate bias.}
    \label{fig:time-res}
\end{figure}

For the measurements of time resolution of the MALTA sensor, we add a dedicated TDC to the readout system, which digitises the arrival time of the MALTA digital output signal. We use the PicoTDC\footnote{CERN PicoTDC ASIC Project https://espace.cern.ch/PicoTDC}, which registers the MALTA signal with 12~ps accuracy. The time resolution is measured in the beam telescope setup as described above with a total of three planes, two reference planes are based on epitaxial MALTA sensors (denoted as $P_{0}$  and $P_{2}$ ) and a DUT sensor in its middle (either Czochralski MALTA sensor or epitiaxial MALTA sensor $P_{DUT}$). The time resolution for each plane can be determined through the measurement of time differences between the combination of planes,
\begin{equation}
\sigma_{12}^{2} = \sigma_{1}^{2} + \sigma_{2}^{2},~~~  \sigma_{13}^{2} = \sigma_{1}^{2} + \sigma_{3}^{2},~~~ \sigma_{23}^{2} = \sigma_{2}^{2} + \sigma_{3}^{2} \\
\end{equation}
\begin{equation}
\sigma_{2} =\sqrt{\frac{\sigma_{12}^{2} + \sigma_{23}^{2} - \sigma_{13}^{2}}{2}}  \label{math:tres} \\
\end{equation}
where $\sigma_{1}$,  $\sigma_{2}$ and  $\sigma_{3}$ denote the time resolution of each sensor plane, and $\sigma_{12}^{2}$, $\sigma_{13}^{2}$ and $\sigma_{23}^{2}$ are extracted from the time difference distributions through Gaussian fits to the distributions. The time resolution of the MALTA DUT sensor is given by $\sigma_{2}$, and $\sigma_{1,3}$ are the time resolutions of the MALTA reference sensors on epitaxial substrates. Figure~\ref{fig:time-res}a shows the time difference distribution for the fastest signal arriving from two epitaxial MALTA sensors with continuous n$^-$ layer.  In a beam test using 4~GeV electrons, the substrate voltage on the MALTA DUT Czochralski sensor is varied, whereas the epitaxial MALTA sensors are operated at constant $V_{sub} = -6$~V.  A correction for the signal propagation down the column height has not been applied and approximately 50\% of the sensor height has been illuminated by the beam. The red curve shows the Gaussian fit to the distribution core part. From this fit the time resolution of the Czochralski sensor is extracted, which is shown in figure~\ref{fig:time-res}b as a function of substrate bias. The core time distribution yields a time resolution of less than 2~ns for high substrate voltages.

\begin{figure}
    \centering
    \includegraphics[width=.9\textwidth]{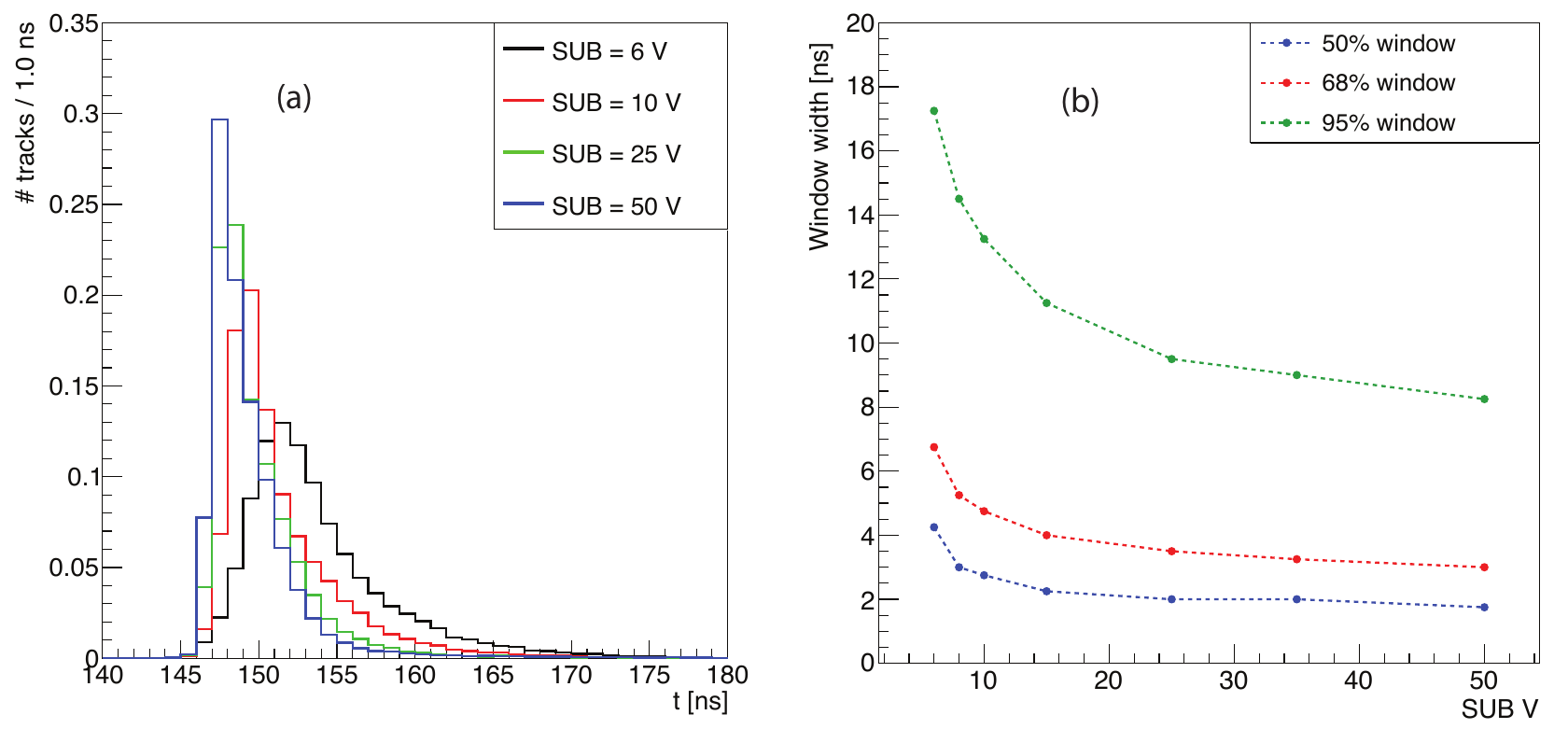}
    \caption{Time difference between Czochralski-MALTA sensor and trigger scintillator at different substrate bias voltages (a). Plot (b) shows the 50\%-, 68\%- and 95\%-integral of distributions in (a) obtained at different substrate voltages. The MALTA sensor is an unirradiated sensor with continuous n$^-$ layer.}
    \label{fig:time-window}
\end{figure}

The time resolution is further investigated using the telescope in the laboratory with a $^{90}$Sr radioactive source and using cosmic rays, where the track is reconstructed in three adjacent planes as described in section \ref{sec:eff}. The time difference between the fastest MALTA signal and the trigger scintillator is measured using the PicoTDC.  Figure~\ref{fig:time-window}a shows the time-difference distributions for the  
Czochralski-MALTA sensor with continuous n$^-$ layer and the trigger scintillator at different sensor bias voltages for hits that are matched to the reconstructed telescope tracks.  No correction is applied for the time resolution of the scintillator ($\sigma \approx$0.5~ns) or trigger logic ($\sigma \approx$0.9~ns). The plot clearly shows how the faster signal and higher amplitude at large substrate voltages for Cz substrates reduces time-walk and narrows the time-difference distribution. 

For applications where the sensor signals need to be registered within a clock-cycle, e.g. the 25~ns LHC bunch-crossing clock, this improvement substantially improves the ``in-time'' efficiency, i.e. the efficiency of detecting the signal in the correct bunch-crossing. To illustrate this we plot the 50\%-, 68\%- and 95\%-integral of the time-difference distributions as function of substrate bias in figure~\ref{fig:time-window}b. The figure indicates that MALTA sensors produced on Czochralski substrate are capable of full in-time efficiency ($>$95\%) up to bunch crossing rates of 100~MHz.

\section{Conclusions}

With the MALTA monolithic CMOS sensor we have investigated design and processing optimisation for high-granularity pixel sensors with excellent radiation hardness, position and time-resolution. With the presented combination of pixel design and sensor processing we achieve full radiation hardness to a fluence of $>10^{15}$ n$_{eq}$/cm$^{2}$ in a full-size monolithic CMOS sensor with small electrodes. Measurements on full-size sensors have also shown excellent time resolution of $<$2~ns in a 512$\times$512 pixel matrix of 36.4$\mu$m pixel pitch. We have produced the identical MALTA sensor design on high-resistivity epitaxial and Czochralski substrates to allow a direct performance comparison between the two substrate choices. The high bias voltages achieved on sensors with high-resistivity Czochralski substrates enables a large depletion volume which results in a significantly larger ionization charge signal in Czochralski MALTA sensors than epitaxial MALTA sensors. This signal gain leads to a superior efficiency after irradiation. Our measurements yielded an efficiency of  $>95$\% for $2\times10^{15}$ n$_{eq}$/cm$^{2}$ irradiated Czochralski MALTA sensors. In a direct comparison of unirradiated expitaxial and Czochralski MALTA sensors we observe significantly larger cluster width on Czochralski sensors, explainable by larger depletion depth, which can be exploited for charge-interpolation to improve spatial resolution in tracking detectors.

\section{Acknowledgements}
We are grateful to the Institute Jo\v{z}ef Stefan, Ljubljana, Slovenia, for their support during neutron irradiations. The irradiation campaign has been supported by the H2020 project AIDA-2020, GA no. 654168. This research project has received funding by the Marie Sklodowska-Curie Innovative Training Network of the European Commission Horizon 2020 Programme under contract number 675587 ``STREAM''.

\bibliographystyle{JHEP}
\bibliography{Malta-Cz}

\end{document}